\documentclass[a4paper]{article}
\usepackage{a4wide}
\usepackage{graphicx}
\usepackage[T1]{fontenc}
\usepackage[utf8]{inputenc}
\usepackage[UKenglish]{babel}
\usepackage{siunitx}
\usepackage{color}
\usepackage{amsmath}
\usepackage{amssymb}
\usepackage{mathtools}
\usepackage[backend=bibtex,style=ieee,citestyle=numeric-comp]{biblatex}
\usepackage[hyperindex,breaklinks]{hyperref}

\graphicspath{{figures/}}

\newcommand{\bonn}{
	\\\textit{\footnotesize Helmholtz-Institut f\"{u}r Strahlen- und Kernphysik,
	Rheinische Friedrich-Wilhelms-Universit\"{a}t Bonn, Germany}
}

\begin{document}
	\title{Physics of Beer Tapping -- Lower vs. Upper Bottle}
	\author{Johann Ostmeyer\bonn}
	\date{\today}
	\maketitle
	\begin{abstract}
		``Beer tapping'' is a well known prank where a bottle of carbonised liquid strikes another bottle of carbonised liquid from above, with the usual result that the lower bottle foams over whereas the upper one does not. Though the physics leading to the foaming process in the lower bottle has been investigated and well documented, no explanation to date has been provided why the upper bottle produces little to no foam.  In this article we describe the reasons for the entirely different behaviours of the two bottles.
	\end{abstract}

	\section{Introduction}
	Recently it was shown in~\cite{beer_tapping,Manti_Lugo_2015} why a bottle of supersaturated carbonised liquid foams over, after being struck from above by another bottle, usually also containing supersaturated liquid. I.e.\@ the gas in the liquid expands and the ensuing foam rises and spills over the container. The beer-drinking community is well versed in such a phenomenon. Such an event, usually performed as a prank, is called `beer tapping'. Here the victim of the prank holds an open beer bottle, while a prankster strikes the top of the beer bottle with the bottom of their own (open) beer bottle. The beer in the lower bottle then foams over the opening and onto the victim's hands, hopefully eliciting some good-natured laughs. The physics dictating the foaming of the beer is well described in~\cite{beer_tapping} and we provide a cursory explanation in \autoref{sec:phys_foam} closely following the aforementioned article.
	
	Missing from~\cite{beer_tapping,Manti_Lugo_2015} is an explanation for why the prankster leaves the scene relatively unscathed and dry. They also hold an open beer bottle, which they use to strike the victim's bottle, yet little to no foam exits their bottle. This outcome is also easily described using the methods of~\cite{beer_tapping}, and is the subject of this article.
	
	\section{The physics of foaming}\label{sec:phys_foam}
	The most important stages of beer tapping are as follows. The initial shock due to the ``tap'' induces oscillations of the radius $R$ of small bubbles already existing in the liquid. These oscillations are driven by the shock pressure $p_S(t)$ and follow, to a good approximation, the Rayleigh-Plesset-equation~\cite{rayleigh_collapse}
	\begin{align}
	\rho R \ddot{R} + \frac 32 \rho \dot{R}^2 -
	\left(p_0+\frac{2\sigma}{R_0}\right)\left(\frac{R_0}{R}\right)^{3\gamma}
	+ \frac{2\sigma}{R} + \frac{4\mu \dot{R}}{R} &= -\left(p_0 +
	p_S(t)\right)\label{eqn:rayleigh_plesset} 
	\end{align}
	where $R_0\approx\SI{180}{\micro\meter}$ is the typical size of the initial radius of the bubble,
	$\rho=\SI{e3}{\kg/\meter\cubed}$ the density of water,
	$p_0=\SI{e5}{\pascal}$ the ambient pressure,
	$\sigma=\SI{0.0434}{\newton/\meter}$ the surface tension between
	CO$_2$ and water, $\gamma=\num{1.304}$ the heat capacity ratio
	of CO$_2$ and $\mu=\SI{e-3}{\kg/\meter/\second}$ the liquid viscosity.
	
	When reaching the first minimum in radial size after $t_\text{min}\approx\SI{0.15}{\milli\second}$ the bubbles in the lower bottle collapse~\cite{hit_the_liquid} according to the model of Brennen~\cite{brennen_2002} into about $N$ fragments of nearly equal size, where
	\begin{align}
	N&=\left(\frac 13 \left(\sqrt{7+3\Gamma}-2\right)\right)^3\,,\\
	\Gamma&=\frac \rho\sigma R(t_\mathrm{min})^2\ddot{R}(t_\mathrm{min})\ .
	\end{align}
	Each initial bubble, therefore, forms a cloud of $N$ smaller bubbles, where typically $N\sim\num{e6}$.
	A thus increased number of bubbles leads to a larger interface area between gas and liquid. This leads to the second stage where the gas within the liquid diffuses into the bubbles of each cloud. The volume of the cloud formed by the bubbles grows in square-root time $\sqrt{t}$~\cite{square_root_growth} for about the next $\SI{10}{\milli\second}$.
	
	Finally, during the last stage, the cloud of bubbles becomes so large that it quickly rises up because of its buoyancy. As it moves, it collects even more gas, which in turn further increases the sizes of the bubbles. Cloud growth now scales as $t^2$ and the cloud takes a form reminiscent of a mushroom cloud from a nuclear explosion (see the illuminating figures in~\cite{summary_beer_effects}). 
	If enough large clouds have been formed, the liquid foams over at approximately $\SI{0.5}{\second}$ after the ``tap''.
	
	The interested reader may find a more detailed physical explanation of the foaming process, as well as many interesting phenomena appearing in carbonised beverages, in reference~\cite{summary_beer_effects}.
	
	\section{Comparison of upper and lower bottle}
	The last two stages do neither depend on the bottle position, nor on the position of a bubble in the bottle. Thus it is sufficient to investigate the collapse of a single bubble at a fixed position in order to understand why the upper bottle (which we now label as U) does not react as vehemently as the lower bottle (now labelled L). Indeed, it is this inequality between the two bottles which makes the prank attractive in the first place.
	
	\subsection{Initial pressure}
	To do so we first have to understand the time evolution of the shock wave. %Let us consider a bubble in a fixed position in space. 
	The pressure induced by the ``tap'' can be modelled by a damped oscillation
	\begin{equation}\label{eqn:damping}
	p_S(t)=p_A\sin\left(\frac{2\pi t}{T}\right)\exp\left(-\frac{t}{\tau}\right)
	\end{equation}
	where $p_A$ is the initial amplitude set by the strength of the ``tap'', $T=\SI{0.24}{\milli\second}$~\cite{beer_tapping}
	and $t$ represents the time since the shock first reached the bubble. It turns out that the shock wave will lose most of its intensity through the reflection at the fluid's surface (implying that the damping does not behave as smoothly as modelled in reality) and not while travelling in the fluid~\cite{beer_tapping}. Thus we expect a damping in the order $\tau\approx 2T$, i.e.\@ the time needed to reach the surface and be reflected.
	However we will find that the damping does not have significant influence on the qualitative behaviour of the foaming.

\begin{figure}[ht]
		\centering
		\input{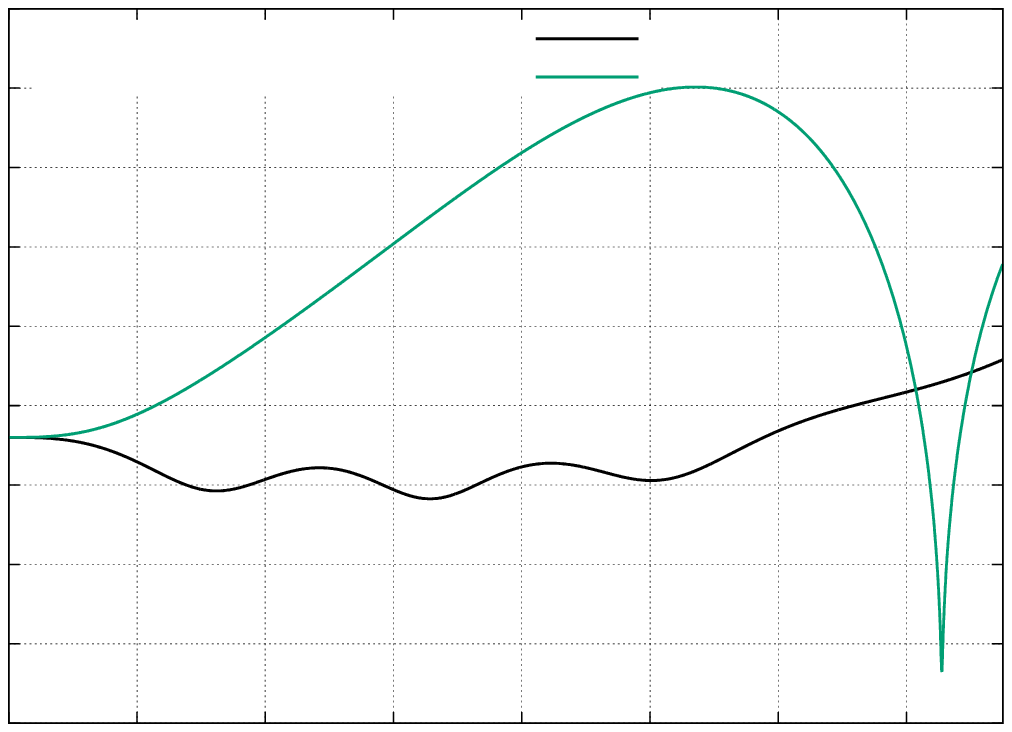}
		\caption{Radius of a CO$_2$-bubble since the first excitation without damping, $\tau=\infty$. Obtained via numerical simulations using Runge-Kutta 4. The black (green) curve shows the behaviour of a bubble in the upper (lower) bottle exposed to positive (negative) initial pressure by the ``tap''.}
		\label{fig:blasenradius}
	\end{figure}

	The difference between bottles U and L now is only the sign of $p_A$, where $p_A>0$ in U and $p_A<0$ in L. This means that bubbles in bottle U are compressed by the shock and oscillate moderately, whereas bubbles in bottle L first expand and then collapse violently. This can be observed in figure~\ref{fig:blasenradius} where we plotted the time evolution of equation~\eqref{eqn:rayleigh_plesset}. The solution of this ordinary differential equation is numerically not challenging and has been performed by the classical Runge-Kutta 4 method. Measurements performed in~\cite{beer_tapping,Manti_Lugo_2015} show good agreement with the numerical predictions for different initial conditions.
	
	\subsection{Bubble collapse}
	The collapse of the oscillating bubbles is described by the model of Brennen~\cite{brennen_2002} according to which the most unstable mode
	\begin{equation}
	n_m=\frac 13 \left(\sqrt{7+3\Gamma}-2\right)
	\end{equation}
	is mainly responsible for the bubble collapse at the first local minimum into $N\approx n_m^3$ fragments. At the same time, $n_m$ gives the factor by which the total surface area of all fragments surpasses the surface area of the original bubble. As this area is proportional to the amount of CO$_2$ entering the bubbles by diffusion, $n_m$ is also a direct indicator for the intensity of the foam formation.
	
\begin{figure}[ht]
		\centering
		\input{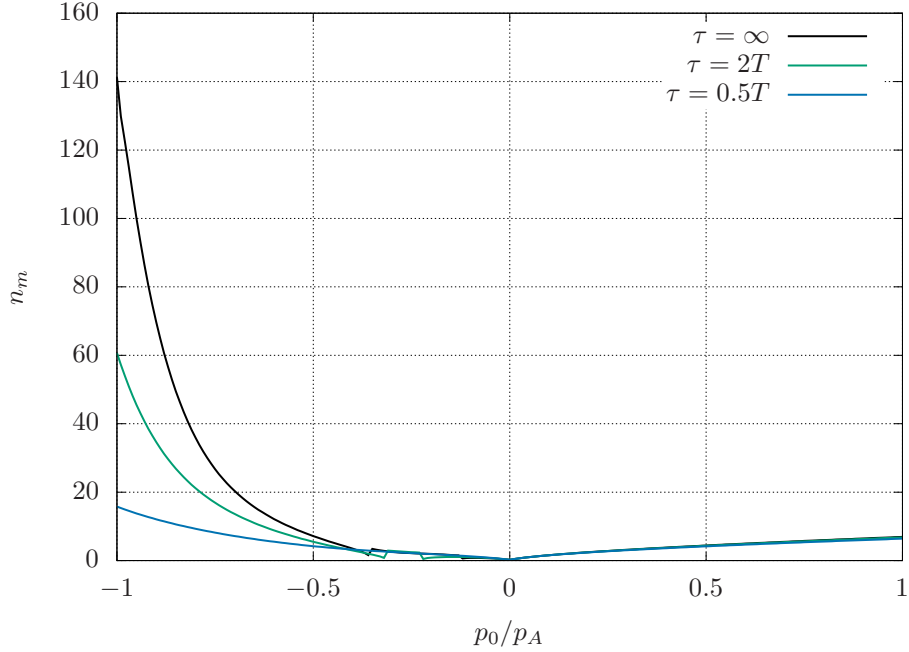}
		\caption{Most unstable mode according to the model of Brennen depending on the initial pressure. The different lines correspond to different damping strengths in eq.~\eqref{eqn:damping}. The black line has been calculated in the absence of damping.}
		\label{fig:unstable_mode}
	\end{figure}

	We calculated $n_m$ for different initial pressures and damping strengths. The results can be found in figure~\ref{fig:unstable_mode}. We observe that the most unstable mode grows in both bottles with increasing hit strength.\footnote{The discontinuities in the region $-0.5<p_0/p_A<0$ are not to be taken to seriously. Here the first local minimum disappears, so that another one is the first, and later appears again. There is no relevant physics hidden in this strange behaviour.} However, as long as the damping is not extremely large, $n_m$ grows very fast for decreasing $p_A$ in bottle L ($-p_0<p_A<0$, the lower bound is needed because no pressure below zero, i.e.\@ vacuum, is possible), but only slowly for increasing $p_A$ in bottle U ($0<p_A$). In~\cite{beer_tapping} $n_m\approx 100$ is observed. This means that the damping is quite small in reality and $\tau\gtrsim 2T$.
	
	We cannot determine the damping quantitatively, but we do not have to, either. It is sufficient to observe that for any realistic damping, $n_m$ can be an order of magnitude larger in bottle L than in bottle U. The only requirement for this to be the case is such a strong ``tapping'' that $p_A\approx -p_0$.

	Obviously the pressure in bottle U is not limited from above, so an extremely strong hit can produce $p_A\gg p_0$ and cause the upper bottle to foam over as well. But such a brute force ``tap'' is not of high interest and can incur damage to the bottles.
	
	\section{Summary}
	Although the precise physics of beer tapping including fluid dynamics are quite complicated, it can easily be understood why the lower bottle L foams over while the upper one U (usually) does not. The ``tap'' creates a low pressure in bottle L which causes existing bubbles of CO$_2$ to expand and then collapse into many fragments. The subsequent CO$_2$-liquid interface grows and this in turn causes more gas to diffuse into the bubble cluster. It rises upwards and creates the foam. In bottle U the high initial pressure induces only moderate bubble oscillations such that the collapse does not happen or, if it does, it does so not as violently. Either way, there is no rapid cloud growth and the prankster leaves the scene dry.
	
	\section*{Acknowledgements}
	Huge thanks to Thomas Luu for proofreading and giving a lot of helpful comments. I would also like to thank Kathrin Grunthal, Carsten Urbach and every other person I ever drank beer with for the inspiration and motivation to write this article. 
		
	%\clearpage
	\printbibliography
\end{document}